\documentclass[amssymb,aps,showpacs, twocolumn,pre]{revtex4-2}

\usepackage{lineno,hyperref}
\usepackage{footnote}
\usepackage{graphicx}
\usepackage{amsmath}
\usepackage{float}
\usepackage{xcolor}
\usepackage{array}  % for "\extrarowheight" macro
\usepackage[utf8]{inputenc}
\usepackage{nomencl}

\makenomenclature
\begin{document}

\title{Modeling the diffusion-erosion crossover dynamics in drug release}

\author{M\'arcio Sampaio Gomes-Filho$^1$}
\author{Fernando Albuquerque Oliveira$^{2,3}$} 
\author{Marco Aur\'elio Alves Barbosa$^4$}

\affiliation{$^1$Centro de Ci\^encias Naturais e Humanas, Universidade Federal do ABC, 09210-580, Santo Andr\'e, S\~ao Paulo, Brazil.}

\affiliation{$^{2}$Instituto de F\'isica, Universidade de Bras\'ilia, Bras\'ilia-DF, Brazil.}

\affiliation{$^{3}$Instituto de F\'{i}sica, Universidade Federal da Bahia, Campus Universit\'{a}rio da Federa\c{c}\~{a}o, Rua Bar\~{a}o de Jeremoabo s/n, 40170-115, Salvador-BA, Brazil.}

\affiliation{$^{4}$Faculdade UnB Planaltina, Universidade de Bras\'ilia, Planaltina-DF, Brazil.}

% \author{M\'arcio Sampaio Gomes-Filho}
% \address{Centro de Ci\^encias Naturais e Humanas, Universidade Federal do ABC, 09210-580, Santo Andr\'e, S\~ao Paulo, Brazil.}

% \author{Fernando Albuquerque Oliveira}
% % \ead{fao@fis.unb.br}
% \address{Instituto de F\'isica, Universidade de Bras\'ilia, Bras\'ilia-DF, Brazil.}

% \address{Instituto de F\'{i}sica, Universidade Federal da Bahia, Campus Universit\'{a}rio da Federa\c{c}\~{a}o, Rua Bar\~{a}o de Jeremoabo s/n, 40170-115, Salvador-BA, Brazil.}

% \author{Marco Aur\'elio Alves Barbosa}
% % \ead{aureliobarbosa@unb.br}
% \address{Faculdade UnB Planaltina, Universidade de Bras\'ilia, Planaltina-DF, Brazil.}

\begin{abstract}
A computational model is proposed to investigate drug delivery systems in which erosion and diffusion mechanisms are participating in the drug release process. Our approach allowed us to analytically estimate the crossover point between those mechanisms through the value of the parameter $b$ ($b_c = 1$) and the scaling behavior of parameter $\tau$ on the Weibull function, $\exp[-(t/\tau)^b]$, used to adjust drug release data in pharmaceutical literature. 
Numerical investigations on the size dependence of the characteristic release time $\tau$ found it to satisfy either linear or quadratic scaling relations on either erosive or diffusive regimes. Along the crossover, the characteristic time scales with the average coefficient observed on the extreme regimes ({\it i.e.}, $\tau \sim L^{3/2}$), and we show that this result can be derived analytically by assuming an Arrhenius relation for the diffusion coefficient inside the capsule. Based on these relations, a phenomenological expression for the characteristic release in terms of size $L$ and erosion rate $\kappa$ is proposed, which can be useful for predicting the crossover erosion rate $\kappa_c$. We applied this relation to the experimental literature data for the release of acetaminophen immersed in a wax matrix and found them to be consistent with our numerical results.
\end{abstract}

% \begin{keyword}
%     drug release \sep membrane erosion dynamics \sep drug diffusion \sep Weibull distribution
% \end{keyword}

% \date{\today }
\maketitle
%%%+++++++++++++++++++++++++++++++++++++++++++++++++++++++++++++++++++
\section{Introduction}
%%%++++++++++++++++++++++++++++++++++++++++++++++++++++++++++++++++++++

Bioerodible polymers play an important role in the design of pharmaceutical devices in which the drug release mechanism is also determined by polymer erosion in addition to diffusion~\cite{langer1990,brem1990, gopferich1993,brem2001, he2011, kamaly2016, talebian2018, lynch2019}. Biopolymers are commonly used to deliver many kinds of pharmaceutical compounds including as anticancer drugs, contraceptive steroids,  antibiotics, and biomolecules such as proteins and DNA~\cite{ kamaly2016, talebian2018, santos2014, siepmann2011:book, xu2017, Zhang21}. In addition, advances in biodegradable devices with applications for ocular drug delivery systems~\cite{lynch2019, kimura2001} and cancer therapies~\cite{brem1990, brem2001, he2011} have  been reported in the literature. 
In these cases, bioerodible polymers are used for two main purposes: to control the drug release time  and to improve biocompatibility. The latter requirement is of greater importance since in certain therapeutic treatments, as in the case of implantable systems, once the biopolymer has completely eroded and its fragments  absorbed by the body, surgical removal of the implant can be avoided~\cite{santos2014, siepmann2011:book, kimura2001}. Thus, a better understanding of the characteristic release times as well as the erosion mechanism are essential for the development of more effective therapies.

Drug delivery systems based on polymeric materials can control the release rates  by diffusion, erosion, and swelling and the prevalence of one of these processes depends on the interactions between polymer and  the environmental fluid (water or biological fluid), in the sense that the polymeric device may have its physical structure unchanged during the release process or may be subject to a process of swelling or erosion.  When the physical structure of the device remains unchanged, the release is basically controlled by simple diffusion and while erosion (or chemical reactions) of the polymer is taking place release rates can be controlled by the interaction between polymer and fluid molecules. For the case in which the release is controlled by the ability of the polymer to swell, that is, when a fluid penetrates the polymer, it promotes a volume variation in the polymeric matrix, thus generating a strain induced breaking~\cite{Oliveira2000,Maroja01,Dias05,Engelsberg13, Barros19}. In this latter case drug release is controlled by the polymer chain relaxation rate (polymer swelling). In addition, release rates can be controlled by a combination of more than one mechanism and the predominance of one or other depends essentially on the characteristics of the polymer, fluid and physicochemical properties of the drug~\cite{langer1990,fan1989:book}. Many other physical and chemical factors  can also play an important role in this process,  for example,  microparticle size, device geometry and drug dissolution (change of phase), among others~\cite{kamaly2016, talebian2018, xu2017, siepmann2011:book,   siepmann2004, klose2008,siepmann2001a, fredenberg2011}.

In this scenario, the use of computational or mathematical models to describe the release kinetics plays a fundamental role in the development of new pharmaceutical devices. Among some of the benefits, modeling allows one to obtain physical insights into the release mechanism and to optimize an existing drug delivery system, eventually reducing the number of experimental tests and the average time and costs for the production of a new drug release device~\cite{Licata08, siepmann08:IJPharm, siepmann2012,Rahman18, caccavo2019,Pontrelli20, kara20}. 

Erosion dynamics of polymeric materials and its effects on release kinetics have been the subject of many studies; for reviews see~\cite{kamaly2016,xu2017,mircioiu2019}.
Despite all the complexity involved in the release process, theories based on the classical diffusion equation give reasonable results when the main mechanism of drug transport is diffusion. To include the effect of polymer erosion dynamics, it is common to consider the diffusion coefficient as a function of time, position, concentration and, eventually,  porosity, and in some cases it is possible to develop theoretical frameworks that are well in agreement with experimental results~\cite{siepmann2004, klose2008, siepmann2001a, charlier2000, agata2011,Nitanai2012, lao2011, siepmann2002}.  There are also sophisticated theoretical models that account for diffusion, erosion theories,  and many other factors, based on partial differential equations or even coupled models between partial differential equations and Monte Carlo (MC) simulations~\cite{gopferich1995, siepmann2002}. In contrast, only a few microscopic models based on cellular automata or Monte Carlo approaches have been developed to investigate the effects of the polymer erosion and drug diffusion on the context of  the drug release process~\cite{gopferich1997, zygourakis1990, zygourakis1996, bertrand2007, barat06a:SMPT, bezbradica2020}.

In previous works from our group, a lattice gas model was used to simulate drug delivery devices and the effects of the membrane porosity on the drug release process were investigated using the MC approach for capsules in two and three dimensions with different sizes. Scaling relations between  release parameters and porosity were obtained that could be used to fit up to $90\%$ of the membrane content covering the capsule~\cite{gomesfilho2016,gomesfilho2020}. Considering the relevance of bioerodible membranes for the development of new pharmaceutical devices, in this work we generalize our previous model including the membrane erosion dynamics (or pore formation). For the purpose of implementing the erosion dynamics, we assume that on each MC step, there is a probability that a pore (leaking site) will be formed due to a possible reaction between the membrane and the fluid environment particles ({\it e.g.}, water molecules). 

 The first step in achieving the description of any dynamical system is to understand its relaxation mechanism which, for many processes, is related to a nonexponential behavior~\cite{Lee83,Morgado02,oliveira19, Vainstein06,Vainstein06b, Costa06}. In the case of drug release patterns the nonexponential relaxation processes can be inferred by the fact that the Weibull distribution function is commonly used to adjust, both in experimental setups and Monte Carlo simulations, release data. Based on previous work, we adopt this distribution in the form~\cite{weibull1951, weibull1952:discussion,  Langenbucher72, slater09}
 \begin{equation} \label{eq:weibull}
  \frac{N(t)}{N_{0}} = \exp \left [- \left ( \frac{t}{\tau} \right )^{b} \right],
 \end{equation}
 where $N(t)$ is the amount of drug particles inside the capsule as a function of time $t$, while $N_{0} = N(0)$ is the initial number of particles; the characteristic release time $\tau$ is associated with the time in which $\approx 63\%$ of the drug is delivered, whereas the release parameter $b$, in the pharmaceutical literature, is associated with the physical mechanisms that control the release process (for more details, see~\cite{gomesfilho2016,gomesfilho2020} and references therein), which has been successfully applied in different studies~\cite{Mohammadpour21,Corsaro21,Berthault21}.  It is important to mention that Ignacio and Slater proposed a new semiempirical function based on diffusion theory  that for a purely diffusive drug release problem outperforms the Weibull function and, recently, they proposed an alternative way to obtain the diffusion constant from the drug release data~\cite{ignacio2017, ignacio2021}.
 
In our simulations, the Weibull distribution function was found to reasonably adjust the release curve for different erosion rates and we were able to perform a scaling analysis on the parameters $b$ and $\tau$ as a function of the erosion rate and the capsule size. By comparing the Weibull parameter $b$ against the erosion rate it was possible to identify a crossover region  where the effects of membrane erosion and drug diffusion were contributing with the same weight to the release mechanism. This crossover was identified to occur at $b=b_c \approx 1$, corroborating the discernment of the drug release mechanism proposed by Papadopoulou~\textit{et al.}~\cite{papadopoulou2006}. A simple mathematical argument was presented to explain this behavior, favoring the usage of the Weibull parameter $b$ to distinguish between the major drug release mechanisms~\cite{papadopoulou2006}.

In addition, the characteristic release time $\tau$ was found to satisfy a power law dependence in terms of the membrane erosion rate and this allowed us to build an analytical expression relating those two quantities. This relation was found to be consistent with experimental data for the release of acetaminophen immersed in a wax matrix~\cite{agata2011}.

The remainder of this paper is organized as follows: in Sec.~\ref{sec:model} we introduce our lattice model and the simulation methods used to investigate the influence of the membrane erosion dynamics on the drug release mechanism. Results and discussions are presented in Sec.~\ref{sec:results} while our conclusions are summarized in Sec.~\ref{sec:conclusion}.

%%+++++++++++++++++++++++++++++++++++++++++++++++++++++++++++++++++++++++++++++++++++++++++++++++++++++++++++++++++++++++++++++++++++++++++++++
\section{\label{sec:model}Model and Simulation Details}
%%%+++++++++++++++++++++++++++++++++++++++++++++++++++++++++++++++++++++++++++++++++++++++++++++++++++++++++++++++++++++++++++++++++++++++++++++
To investigate how the membrane erosion influences the release kinetics, we modified the two-dimensional (2D) lattice model of drug release~\cite{gomesfilho2020, gomesfilho2016}. Instead of presenting a rigid membrane with fixed porosity, with the random porous sites distributed along the membrane at the beginning of the simulation, we encapsulated the device with a bioerodible membrane, whose dynamics is defined along the simulation, as described below.
 
\begin{figure}[tbp]
	\begin{center}
 	  \includegraphics[width=1\columnwidth]{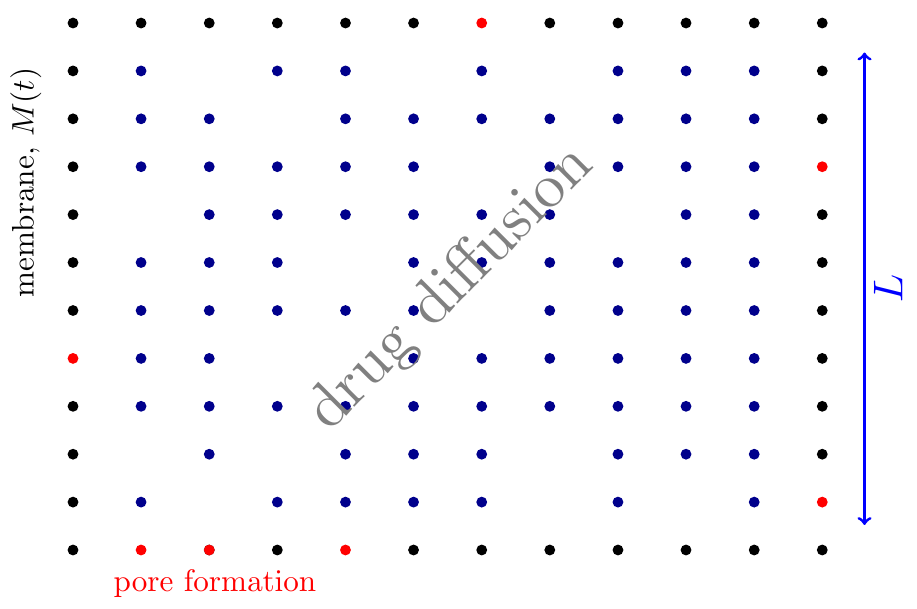}	
	\caption{\label{model} Schematic representation of our two-dimensional device model with size $L=10$, $N_0 = L^2$, and $M_0=4L$ for a given time step. The lattice is not represented, for more clarity. Nonetheless, each site has area of $l_0^2$ and can be occupied by a single  particle or be empty. The erosion constant dictates the formation of pores, with six  membrane sites (black)  eroded (red) and some of the drug particles (blue) released. }
	\end{center}
\end{figure}

In our model, the drug device is represented by a lattice and each site can be occupied by a single drug particle or be empty.  Drug dynamics occurs in a random way, but drug molecules are not allowed to jump into other sites occupied by either drug or membrane particles, \textit{i.e.}, the main effect of the membrane is to block drug molecules from leaking to the outside environment. Except for membrane dynamics, simulations are similar to our previous work and we refer to it for further details~\cite{gomesfilho2020}. We start all simulations by considering 2D square lattices with size $L$, maximum initial drug concentration ($N_0 = L^2$), and membrane coverage ($M_0=4L$).  A pictorial  representation of our 2D device model is presented in Fig.\ref{model} for a given time step.

The membrane erosion dynamics (or pore formation) is introduced in our lattice model as simple as possible with membrane and drug dynamics being independent of each other along a Monte Carlo  step. Between MC steps there is a constant probability that $\delta M$ pore sites (leaking site) will be formed due to a possible reaction between a membrane site and an environment fluid particle (it is assumed that the system is immersed in an implicit aqueous medium). 

Before proceeding, let us represent the erosion rate constant as $k = P \delta M l_0/\Delta t$, where $P$ is the probability of erosion, $l_0$ is the pore length, $\Delta t$ is the unit of time assumed for a MC step, and  $\delta M$ is the number of pores eroded along the $\Delta t$ time interval. After each MC step, a random number $x$ uniformly distributed between $0$ and $1$ is generated and, if $x \leq P$, $\delta M$ new pores are created in random membrane sites. 

For each configuration, {\it i.e.}, different sizes $L$ and erosion rates $k$, the final drug release profile and membrane coverage are obtained by averaging the particle number and membrane sites, as a function of the time, over $10^3$ simulations. Each individual release curve is then adjusted to the Weibull distribution function using standard routines for non linear square fitting~\cite{gnuplot}.

In Fig.~\ref{model}, we show, for example, a possible obtained configuration for a given time step. As the final results are obtained as an average over different realizations,  and the  constant erosion probability is assumed, the average number of membrane sites decays linearly with time,
\begin{equation} \label{eq:membrane}
M(t) = M_0 - k t.    
\end{equation}
Generalizations of our microscopic probabilistic approach are straightforward to produce different decay profiles, $M(t)$, but here we are interesting in the process where the polymer membrane erosion follows a  linear relation with time~\cite{agata2011, caccavo2019}. Furthermore, the erosion constant $k$ is similar to the experimental erosion constants described in the cellular automaton model for the corrosion of a metal in an environment~\cite{saunier2004} and also in a model for swelling-controlled drug release~\cite{laaksonen2009}. Furthermore, it is important to mention that we are interested in the erosion of polymeric membranes, and other methods can be applied to investigate the erosion of polymeric matrices~\cite{gopferich1997, Reynolds1998, Rothstein2008}.

For simplicity, let us define the dimensionless erosion constant $\kappa = k \Delta t /l_0$ and consider  $\delta M=1$.  As an illustration, for $\kappa = 1.0$,  all attempts are accepted and 100 membrane sites will erode after 100 MC steps, for a large enough membrane, whereas for  $\kappa = 0.01$ on average only one membrane site is converted into a pore site after $100$  MC steps.

%++++++++++++++++++++++++++++++++++++++++++++++++++++++++++++++++++++++++++++++++++++++++++++++++++++++  
\section{\label{sec:results} Results and discussion}
%++++++++++++++++++++++++++++++++++++++++++++++++++++++++++++++++++++++++++++++++++++++++++++++++++++++  

\begin{figure}[tbp]
	\begin{center}
 	  \includegraphics[width=1\columnwidth]{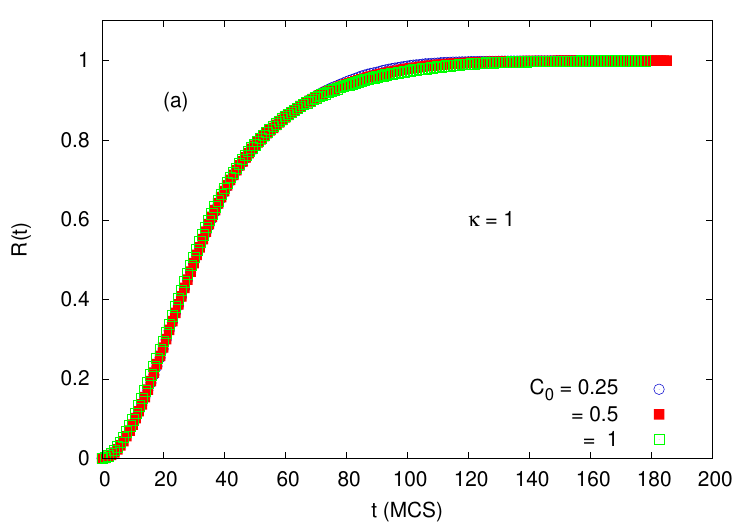}	
 	   \includegraphics[width=1\columnwidth]{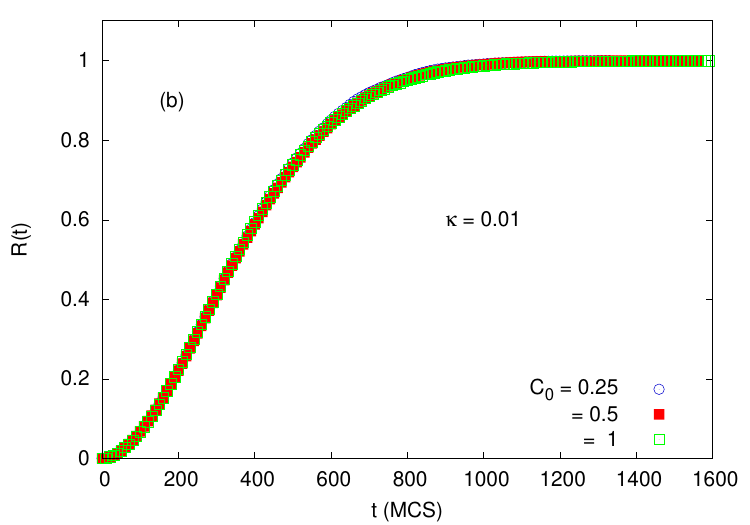}
	\caption{\label{fig1R} Fraction of drug released, $R(t)$,  as a function of the time $t$, in Monte Carlo steps, for  different drug initial concentrations $C_0$ and for erosion constant (a) $\kappa = 1$  and (b) $0.01$. }
	\end{center}
\end{figure}

We start presenting in Fig.~\ref{fig1R} the release fraction $R(t) = 1 - N(t)/N_0$ resulting from  Monte Carlo simulation of drug device models with size $L=200$ and the erosion constant, (a) $\kappa = 1.00$  and (b) $0.01$, for different drug initial concentration, $C_0 = N_0/L^2$, where $N_0$ drug particles are initially placed randomly along the lattice. In this way, these figures indicate that the obtained results are universal since drug release data presents the same functional form dependent only on the erosion constant. On the other hand, this universality is expected to be broken for different devices geometries, for example, for rectangular lattices. Hereafter, we choose  the initial concentration  equal to one, as it seems to be closer to real cases  in the sense that the drug device is completely filled with particles.

\begin{figure}[tbp]
	\begin{center}
 	  \includegraphics[width=1\columnwidth]{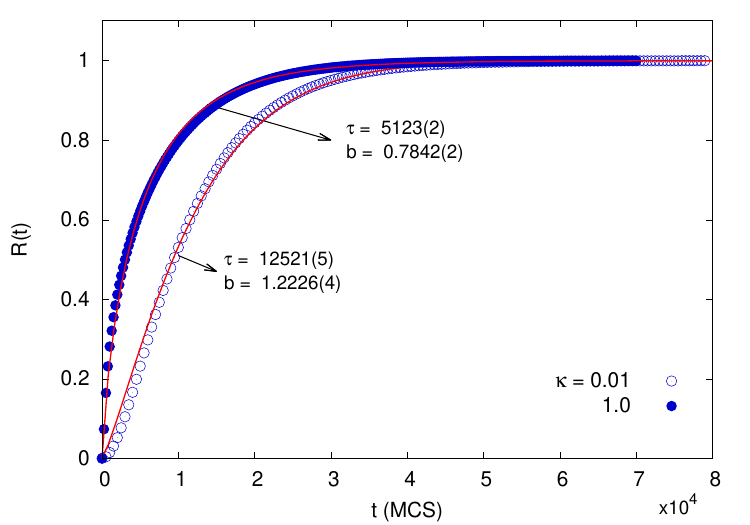}	 
	\caption{\label{fig1}Fraction of drug released  as a function of the time $t$, in Monte Carlo steps, for device models with size $L=200$ and different erosion rates $\kappa$. Simulation data points were adjusted to Weibull distribution functions (lines), with fitting parameters $b$ and $\tau$ indicated along each curve.}
	\end{center}
\end{figure}

In Fig.~\ref{fig1}, we show that the Weibull function provided accurate fits for release curves  for different erosion constant values $ \kappa = 0.01$ and $1.00$, which illustrate the cases where either erosion or diffusion is the main mechanism dominating the drug release pattern. In both cases, the numerical data could be reasonably well adjusted to Weibull functions (continuous lines), as measured by $R^2\approx 1$ and  also through an analysis of its residuals~\cite{gomesfilho2020}.

For $\kappa = 1.00$, the membrane erosion happens so fast that it is almost instantaneous in the time frame presented in Fig.~\ref{fig1}. In this case the Weibull adjusted parameters\footnote{Weibull parameter $b$ is being shown with three significant digits and the error on its value, as obtained from the precision of the adjust, is smaller than the last significant digit.} were found as $\tau= 5123(2)$ MC steps and $b=0.78$, with the latter being used in the literature~\cite{papadopoulou2006} to associate the drug release mechanism with drug diffusion in a regular, Euclidean lattice, but with contributions from another mechanism which, in the case of current model, is immediately recognized as the membrane erosion. The presence of the membrane with a fast erosion dynamics also introduces numerically significant effects on the behavior of the characteristic release time $\tau$, as will be discussed below.

For the slow erosion rate case\footnote{In our model, for a system with $4L = 800$ membrane sites and an erosion rate equal to $\kappa = 0.01$  it would take $8 \times 10^4$ MC steps for the membrane to disappear.}, $\kappa = 0.01$, the membrane degradation takes about twice as much time as what is needed to release $99\%$ of the initial drug load, and the drug release kinetics is mostly limited by the membrane erosion rate. For this case Weibull adjusted parameters were found as $\tau = 12521(5)$ MC steps and $b=1.22$, which could be used to  classify this system as belonging to devices with a ``complex release mechanism''~\cite{papadopoulou2006}, which evidently corresponds to a dominance of the membrane erosion dynamics on the drug release profile. 

As discussed above, the presence of an erodible membrane covering the pharmaceutical device introduces an additional level of complexity on drug release patterns, presenting more important effects than  would be presumed by simply noting that the membrane decreases the probability for drug molecules to escape the capsule. By comparing the cases of a simple device without a membrane to a series of systems covered with sequentially stiffer membranes, the characteristic times would increase but, also, the shape of the release curve would be radically changed, crossing drug release mechanisms from simple diffusion ($0.69 <b < 0.75$), normal diffusion with contribution from another mechanism ($0.75<b<1.0$), first order release ($b=1$) and, finally, to complex release ($b>1$). Since the Weibull distribution function provides a reasonably good approximation to release curves in the cases illustrated in Fig.~\ref{fig1}, it should be interesting to perform a detailed scaling analysis on the dependence of release parameters $b$ and  $\tau$ with respect the capsule sizes $L$ and the erosion rate $\kappa$ within this interval. The following sections will be devoted to this task.

%+++++++++++++++++++++++++++++++++++++++++++++++++++++++++++++++++++++++++++++++++++++++++++++++++++++++++++
\subsection{Relation between release parameter $b$ and erosion rate~$\kappa$}
%+++++++++++++++++++++++++++++++++++++++++++++++++++++++++++++++++++++++++++++++++++++++++++++++++++++++++++
To investigate the interplay between membrane erosion and drug diffusion in the drug release, let us express the membrane fraction at short times ($t \ll \tau $) from Eq.~(\ref{eq:membrane}) as:
\begin{equation}\label{eq:probability-membrane}
    \lambda_M(t) = \frac{M(t)}{M_0} = \frac{M_0 -k t}{M_0} \approx \exp \left (-\frac{k  t}{M_0} \right),
\end{equation}
and compare it to the drug fraction inside the device {\it by assuming} that it can be approximated by the Weibull distribution function in Eq.~(\ref{eq:weibull}),
\begin{equation}
    \lambda_N(t) = \frac{N(t)}{N_0} = \exp \left [- \left ( \frac{t}{\tau} \right )^{b} \right].
\end{equation}

\begin{figure}[tb]
	\begin{center}
 	 \includegraphics[width=1\columnwidth]{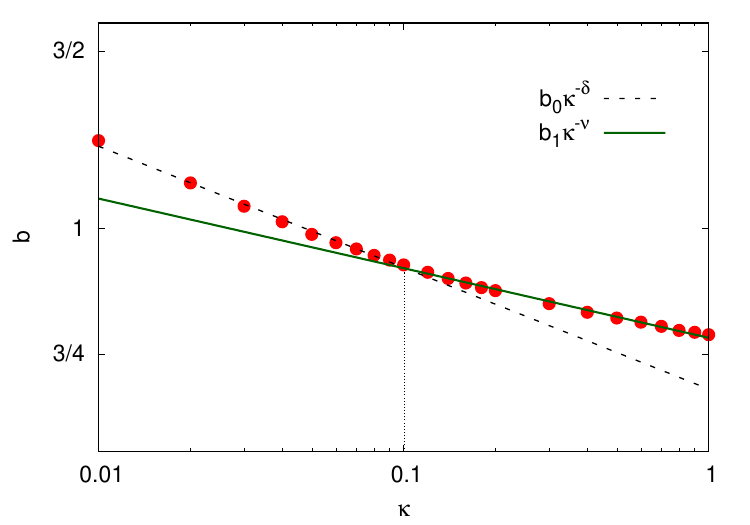}	
	\caption{\label{fig2}  (color online) Log-log plot of the release parameter $b$ as a function of erosion rate $\kappa$ for capsules with size $L=200$. The lines are the fitted curves (see text). Note that there is a subtle deviation from the fitted curves when $ \kappa \to 1$ and $\kappa \to 0$.}
	\end{center}
\end{figure}

Although the membrane kinetics is decoupled from the drug diffusion, both mechanisms are acting together in a non trivial way to generate the final drug release profile. Under the condition that both mechanisms are contributing in a proportional way to the system dynamics (\textit{crossover}), we shall have  a very special case, such that the ratio $\frac{\lambda_M(t)}{\lambda_N(t)}$ is constant and equal to one, which holds true only if $b=1$ and $\tau = M_0/k$, implying that in the crossover region, the drug release is described by an exponential decay. This is in accordance with the experimental classification scheme discussed above~\cite{papadopoulou2006} and it is investigated with more numerical detail in Fig.~\ref{fig2}, which presents the Weibull release parameter $b$ for a capsule with size $L=200$ by varying $\kappa$ from $0.01$ to $1.00$.
% \begin{equation}
%     \dfrac{\kappa t}{M_0} = \left ( \dfrac{t}{\tau}  \right )^b,
%     \label{eq:crossover0}
% \end{equation}
In this figure, presented in log-log scale, two different behaviors seem to occur for the data below and above $\kappa_c \approx 0.1$. In both cases the data are well adjusted by power law scaling relations, but with different coefficients. For $\kappa <  \kappa_c$, one can adjust $b$ data to: $b = b_0\kappa^{-\delta}$,
% \begin{equation} \label{eq:b0}
%  b = b_0\kappa^{-\delta},
% \end{equation}
while in the other limit, $\kappa > \kappa_c$, it can be adjusted to: $b = b_1\kappa^{-\nu}$,
% \begin{equation} \label{eq:b1}
%  b = b_1\kappa^{-\nu},
% \end{equation}
where $b_0$, $b_1$, $\nu$, and $\delta$ are adjustable parameters\footnote{Adjusted parameters are equal to $b_0 = 0.6924\pm0.0006$, $\delta = 0.120\pm0.003$, $b_1 = 0.778\pm0.002$ e $\nu = 0.069\pm0.001$.}. 
It is important to mention that these scaling relations only allow us to find the crossover, when both are equal  (see in Fig.~\ref{fig2}), which means:
\begin{equation}\label{eq:crossover}
 b_c(\kappa_c) = \frac{b_1}{b_0} \kappa_c^{~\delta - \nu},
\end{equation}
and, thus, when $\kappa_c = 0.1$ one obtains  $b_c \approx 1.0$, which deviates by about $8\%$ from the exact numerical value, $b_c = 0.92(1)$.

Therefore,  it is possible to infer that the prevalence of the membrane erosion or diffusion mechanism on the release process determines the values observed on $b$. When the membrane erosion rate is fast (\textit{e.g.}, $\kappa = 1$), the drug diffusion is the dominant mechanism determining drug release. 

As the erosion rate $\kappa$ decreases towards the critical value, $\kappa \to \kappa_c $, there is a corresponding increase in the values of $b$, indicating a greater contribution from the erosion mechanism itself, and the two mechanisms occur in a proportional way at $\kappa = \kappa_c$, where we found $b = b_c \approx 1.0$, in accordance with the phenomenological description presented above. For $\kappa$ values smaller than $\kappa_c$, the membrane erosion becomes a limiting step for the process of drug release and, in this regime, the membrane erosion dynamics becomes the dominant mechanism for controlled drug release.

\begin{figure}[t]
	\begin{center}
  	  \includegraphics[width=1\columnwidth]{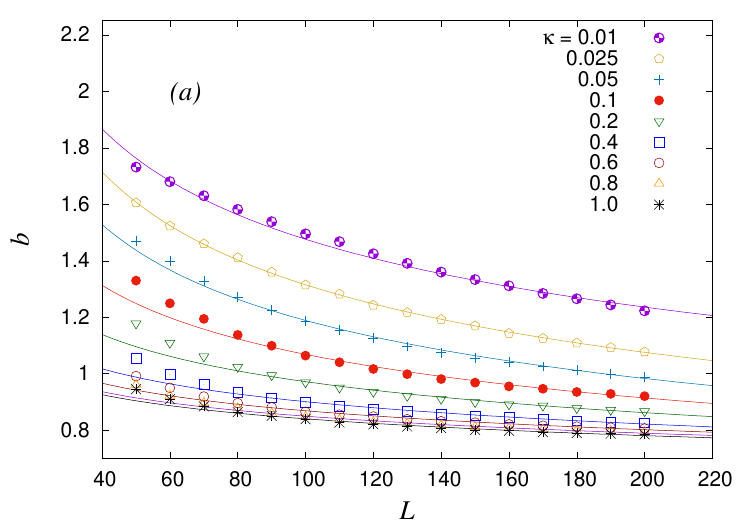}
  	  \includegraphics[width=1\columnwidth]{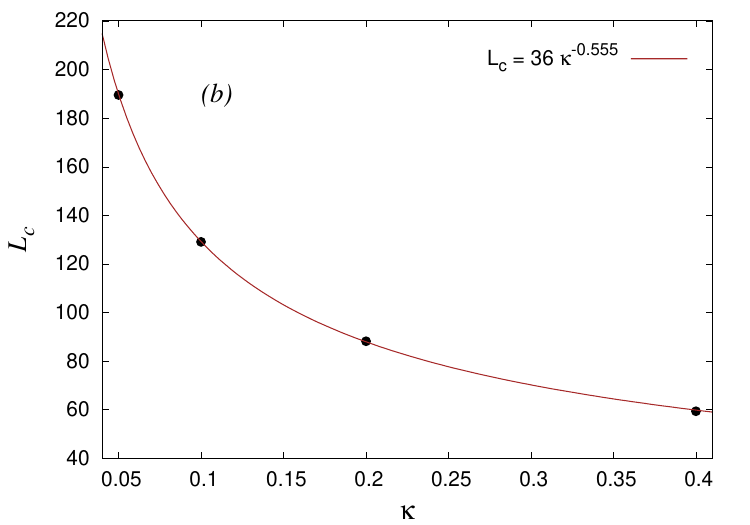}
	\caption{\label{fig3}(a) Release parameter $b$ as a function of the capsule size $L$  for different membrane erosion rates $\kappa$, with the points obtained from the simulation results whereas the lines represent the adjusted curves (see text). (b) The points represent the crossover size $L_c$ against the erosion rate $\kappa$, whereas the line represents the fitted function.}
	\end{center}
\end{figure}

In Fig.~\ref{fig3}(a) the release parameter $b$ is shown as a function of size $L$ for different erosion rates $\kappa$. At fixed values of $\kappa$,  the values of $b$ decrease with increasing size  $L$  in a behavior which can be approximately described by a power law, $b = b_\kappa L^{-\alpha_\kappa}$, where the adjusted parameters $b_\kappa$ and $\alpha_\kappa$ also depend on the erosion rate. Furthermore, the observed size dependence of $b$ allows us to question what the crossover size $L_c$ would be for each $\kappa$, as depicted in Fig.~\ref{fig3} (b). From these two figures, it is evident that there is a threshold size $L_c$ below which membrane erosion starts to be the dominant effect, for each $\kappa$, while above this size the drug diffusion becomes more relevant. These results reinforce the idea that the release parameter $b$ could be used to discern the drug release mechanism, in the sense that, for values $b > 1$, the system is controlled by the  erosion mechanism. 

%+++++++++++++++++++++++++++++++++++++++++++++++++++++++++++++++++++++++++++++++++++++++++++++++++++++++++++
\subsection{Relation between characteristic time $\tau$ and erosion rate $\kappa$\label{sec:tau-kappa}}
%+++++++++++++++++++++++++++++++++++++++++++++++++++++++++++++++++++++++++++++++++++++++++++++++++++++++++++
Next we investigate the size contribution of the capsule to the release parameter $\tau$ for different erosion rates. In particular, this information is useful for understanding the interplay role of membrane erosion and drug diffusion on the drug release process, and its role in creating a crossover region. Before starting, it should be stressed that as we are dealing with an out-of-equilibrium mesoscopic system and that, considering this, the observed crossover regions are not expected to occur concomitantly for the $b$ and $\tau$ release parameters. Nevertheless, while increasing the system size, it should be reasonable to expect that crossover signature in $\tau$ becomes closer to the those obtained for $b$.

Insights on the scaling behavior of $\tau$ with the capsule length can be obtained if one notes that the stochastic diffusion of the drug molecules inside the capsule satisfies: 
\begin{equation}
  \left < r^2 \right > \sim 2d D_0 t
\end{equation}
for long times, with $\left< r^2 \right >$  the drug mean square displacement, $d$ the system dimension, $D_0$ the diffusion coefficient of the drug inside the capsule, and $t$ the time~\cite{Salinas:book, oliveira19}.

For a diffusion controlled devices, where erosion is faster than diffusion ($\kappa \gg \kappa_c $), the characteristic time $\tau$ will be mostly determined by the previous equation, scaling with the capsule size $L$ as $\tau = a L^2$, with $a = 1/(2dD_0)$. On the other hand, for erosion controlled devices, where erosion is much slower than diffusion ($\kappa \ll \kappa_c$), becoming a limiting step for drug release, the characteristic time for drug release $\tau$ is expected to follow the trend presented by the membrane behavior and is expected to scale linearly with the size of the device $L$ ($\tau \sim a L$). 

\begin{figure}[t]
    \begin{center}
        \includegraphics[width=1\columnwidth]{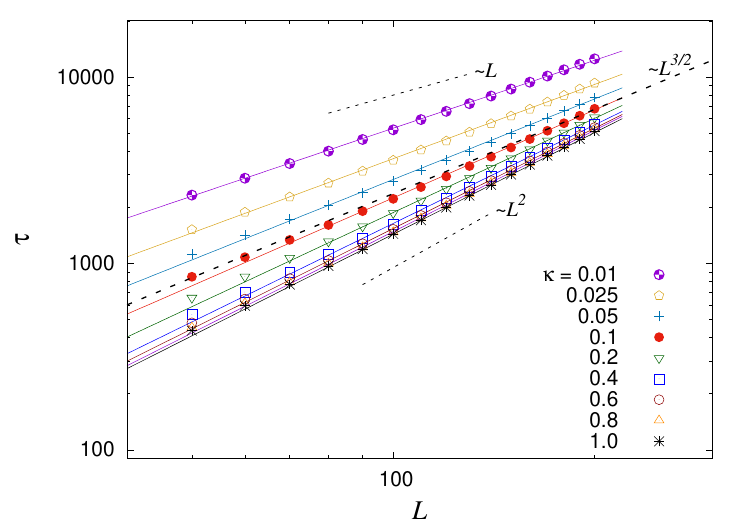}	
        \caption{\label{fig4}(color online) Log-log plot of the characteristic release time $\tau$ as a function of the capsule size $L$  for different membrane erosion rates $\kappa$ with sizes $L$ varying from $50$ to $200$; simulation data marked as points and adjusted curves with lines (see text).}	
    \end{center}
\end{figure}

This pattern is verified numerically in Fig.~\ref{fig4}, where $\tau$ values are shown for sizes between $L=50$ and $200$ and erosion rates $\kappa$ between $0.01$ and $1.00$. It is possible to anticipate how the characteristic time will scale with $L$ in the crossover region by making a simple average of the exponent $\mu$ in the expression $\tau \sim L^{2-\mu}$, considering the extreme cases, resulting in $\mu_c = (0+1)/2 = 1/2$, {\it i.e.}, $\tau \sim L^{3/2}$. Appendix~\ref{append:a}  shows that this scaling relation can be calculated with the assumption that the generalized diffusion exponent for drug molecules inside the capsule satisfies an Arrhenius relation. 

As shown in Fig.~\ref{fig4}, our simulations do indicate that, for large enough $L$, the crossover occur at some critical erosion rate $\kappa_c$ and its signature will be observed both through the scaling relation on $\tau_c \sim L^{3/2}$ and on the observed value of the Weibull parameter $b$ which will be $b_c \approx 1$. In other words, for large enough $L$ the crossover signatures in both $b$ and $\tau$ happen for the same erosion rate $\kappa_c$. Also note that $\tau_c \sim L^{z}$ with $z=3/2$ shows a possible connection with universal growth phenomena~\cite{GomesFilho21}, which 
deserves further investigation. It also makes the electrochemical model~\cite{saunier2004} similar to the etching model~\cite{Gomes19}, which belongs to the Kardar-Parisi-Zhang universality class.

\subsection{An expression for $\tau(L, \kappa)$}

Let us now introduce an expression for $\tau(L, \kappa)$ which adjusts our numerical data and could be useful for extrapolating experimental data and predicting crossover erosion rates $\kappa_c$. Despite being phenomenological by construction, the advantage of the current approach is that it is inspired by reasonable physical arguments regarding the crossover between the two mechanisms present in the current model, and also is consistent with current simulations. Later, we will apply it for the release of acetaminophen from the erosive wax matrix~\cite{agata2011}.

\begin{figure}[ht]
	\begin{center}
  	  \includegraphics[width=1\columnwidth]{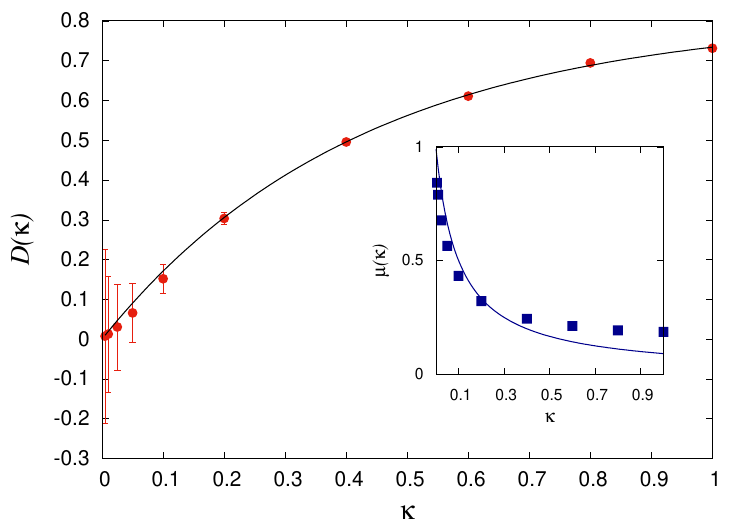} 
	\caption{\label{fig5}(The effective diffusion, $D(k)=D_0 F(\kappa)$,  plotted against the erosion rate constant $\kappa$. Data points correspond to values adjusted from the simulations presented in Fig.~\ref{fig4} and a continuous line corresponds to the function adjusted from~(\ref{eq:deff}). Inset: Similar results for the exponent $\mu(\kappa)$ from~(\ref{eq:muk}).}
	\end{center}
\end{figure}

Considering that in a pure diffusive system $\tau \sim a L^2$, we choose to associate an effective diffusion coefficient of  drug molecules inside the device by defining $a(\kappa) = \frac{1}{2dD(\kappa)}$. Further analysis of the characteristic time presented in Fig.~\ref{fig4} suggests a dependence on the erosion rate $\kappa$ for both the size dependence exponent $\mu \equiv \mu (\kappa )$ and the effective diffusion $D\equiv D(\kappa)$. In this way, we choose to represent
\begin{equation}
   \tau \equiv \tau (L,\kappa)  \approx   \dfrac{l_0^2 L^2}{2dD(\kappa)}  L^{-\mu(\kappa)}. \label{eq:tau-a-mu}
\end{equation}
Guided by the discussion in the previous section, we can approximate the numerical values for $\mu(\kappa)$ with reasonable precision by the function
 \begin{equation}
     \mu (\kappa) = \dfrac{1}{1 +\kappa/\kappa_c}, \label{eq:muk}
 \end{equation}
 which makes $\tau$ satisfy the expected linear (quadratic) behavior in the small (large) erosion limit, with $\kappa_c$ being the size dependent crossover erosion rate, for which $\mu(\kappa_c)=1/2$ (\textit{e.g.}, see Fig.~\ref{fig5}). The effective diffusion $D(\kappa)$ is given by:
\begin{equation}
    D(\kappa) = D_0 F(\kappa), \label{eq:deff}
\end{equation}
with $F(\kappa)$  a function that switches off the diffusion coefficient $D_0$ as the membrane covering the device becomes stiffer. The limiting behavior of $F(\kappa)$ must satisfy
\begin{equation}
    \lim_{\kappa \to 0} F(\kappa) = 0, \nonumber
\end{equation} 
since drug molecules cannot escape from the device without erosion, and 
\begin{equation}
    \lim_{\kappa \to \infty} F(\kappa) = 1,\nonumber
\end{equation}
for recovering the limit of simple diffusion. With these definitions, we were able to use the simulations results presented in Fig.~\ref{fig4} to fit Eq.~(\ref{eq:deff}) with $F(\kappa)$ given by the function:
\begin{equation}
 F(\kappa) = 1 - \exp(-\gamma \kappa), \label{eq:f}
\end{equation}
where $\gamma = 2.36(5)$ and $D_0 = 0.81(1)$ are model dependent constants, as shown in Fig.~\ref{fig5} for the numerical values from simulations and the best fitted functions for both $D(\kappa)$ and $\mu(\kappa)$ (inset).
By using expressions~(\ref{eq:muk})-(\ref{eq:f}) on definition~(\ref{eq:tau-a-mu}), we obtain
\begin{equation}
    \tau (L,\kappa) = \tau_D \dfrac{(l/l_0)^{ -(1 +\kappa/\kappa_c)^{-1}}}{1 - \exp(-\gamma \kappa )} , \label{eq:tau}
\end{equation}
where $l=L l_0$ is the experimental capsule size and $\tau_D$ is a characteristic time that is independent of erosion or other membrane properties, given by
\begin{equation}
    \tau_D = \dfrac{l^2}{2dD_0}.\label{eq:tau_D}
\end{equation}
It should be noted that the product $\gamma \kappa$ is dimensionless and the physical dimension of the diffusion coefficient, $D_0$, is $[\text{length}]^2/[\text{time}]$, which imposes the correct unit of time in the expression~(\ref{eq:tau}).

While expression~(\ref{eq:tau}) was devised in a phenomenological approach to resemble the behavior observed in our model, we note that the main assumptions used in deriving it were related to the comprehension that there is an interplay between two release mechanisms: the erosive mechanism whose dynamics introduces a linear relation between characteristic time and characteristic length, and the random diffusion that drug particles perform {\it inside} the capsule device, which connects characteristic times and  length in a quadratic way. In this way, we expect that expression~(\ref{eq:tau}) could be useful in situations that despite being different from the statistical model used to deduce it, also present those two release mechanisms. This is done in the following section, where we discuss the possibility of applying our approach to the release of acetaminophen (paracetamol) immersed in a wax matrix~\cite{agata2011}. 

 A low erosion rate limit expression for~(\ref{eq:tau}) can be easily calculated, which in turn allows the major physical quantities to be obtained from a simple linear fit. This could be useful for determining both $\kappa_c$ and $\gamma$, but demands data about devices with small erosion rate $\kappa$, as compared to the yet unknown $\kappa_c$, and also small variations in the characteristic time $\tau$ while changing $\kappa$. Due to this requirement, this approach cannot be used with the experimental data discussed in the next section,  but we keep the low-$\kappa$ limit calculation in Appendix~\ref{append:b} as it can be useful in other contexts.

\subsection{Acetaminophen release from a wax matrix}

Agata {\it et al.}~\cite{agata2011} developed a theoretical model based on the solution of the diffusion equation for three dimensional capsules, in which the capsule radius changed with time, $a(t) = a_0 - k t$, with $a_0$ being the initial value of the capsule radius and $\kappa$ is the erosion rate of the eroding front. The model was successfully applied to adjust the release of acetaminophen in a wax matrix in which the erosion was increased with the addition of a pH-dependent functional polymer aminoalkyl methacrylate copolymer E (AMCE) and, besides accurately adjusting release curves, they also obtained values of erosion and drug diffusion rates for 30 experimental batches with different concentrations of acetaminophen and AMCE, as well as different capsule sizes and pH values. The fact that in this work both erosion and diffusion rates can be calculated make it possible to use those data as input in~Eq.~(\ref{eq:tau}) in order to calculate the main new feature pointed by our model: the crossover erosion rate $\kappa_c$. Nevertheless, predicting $k_c$ from the data available on~\cite{agata2011} is challenging because, even after using $k$ and $D_0$ from this paper, Eq.~(\ref{eq:tau}) is still left with three quantities to be determined: $k_c$, $\gamma$, and $l_0$, but on this work at most two batches were prepared with equivalent sizes and fraction of acetaminophen. 

To overcome this difficult we first estimate the pore size $l_0$ using the diameter of the acetaminophen molecule in a spherical molecular approximation, $l_0 \sim V^{1/3}$, which can be calculated as $l_0 = 5.7805 \times 10^{-4}\mu$m ($\text{C}_8\text{H}_9\text{NO}_2$) by using its molecular weight and density values, which are $151.16$ g/mol and $1.3$ g$/\text{cm}^3$, respectively~\cite{PubChem}. With this value  Eq.~(\ref{eq:tau}) is undetermined by two parameters, $\kappa_c$ and $\gamma$, but it is still necessary to investigate the meaning of characteristic time $\tau$ in the experimental setup\footnote{We choose to be careful on the interpretation of $\tau$ in relation to the experimental data since we are working with a phenomenological theory and, considering this, it should be important not to input biases ({\it e.g.}, toward the usage of the Weibull function) in our assumptions.}. 

Within this context we devised two criteria for selecting a batch: $(i)$ the capsules should be big enough in order to reduce small size effects and, particularly, considering that we observed that different signatures for the crossover between erosive and diffusive regimes converge to the same values of erosion for bigger systems, and $(ii)$ the capsule should be eroding as slowly as possible, in order to distinguish it properly from the diffusive regime, which is well described by diffusion equation without erosion. In this way we choose to apply Eq.~(\ref{eq:tau}) to batch $5$ at pH $6.5$, as indicated on Table~$2$ of Ref.~\cite{agata2011}, and we use  their estimates for erosion rate, $k = 6.37\times10^{-2}\mu$m/min, diffusion coefficient, $8.07\times10^{-2} (\mu \text{m})^2$/min and  mean particle radius size, $l=234.1\pm6.9\mu$m. We extracted the drug release data for this batch from Fig.~$1$ of Ref.~\cite{agata2011} and adjusted it to Weibull distribution function, having found $b = 1.12(4)$ and $\tau = 1149(23)$~minutes which, in our proposition, indicates that the erosion mechanism controls the release dynamics. 

 Next, we show that by choosing some $k_c$ (values around the  experimental erosion constant) as a function of $\gamma$, we are able to find the drug release time in which, for some $\{k_c, \gamma\}$, falls in the experimental range. Although this process seems  naive, it can be used by an experimental researcher to make some predictions about the drug release  mechanism and/or improve a particular drug device in order to obtain a desired drug delivery (weeks, months, etc), which is extremely difficult to achieve by  knowing only the type of the  polymer matrix and the diffusion coefficient of the pharmaceutical component.

\begin{figure}[tbp]
	\begin{center}
 	  \includegraphics[width=1\columnwidth]{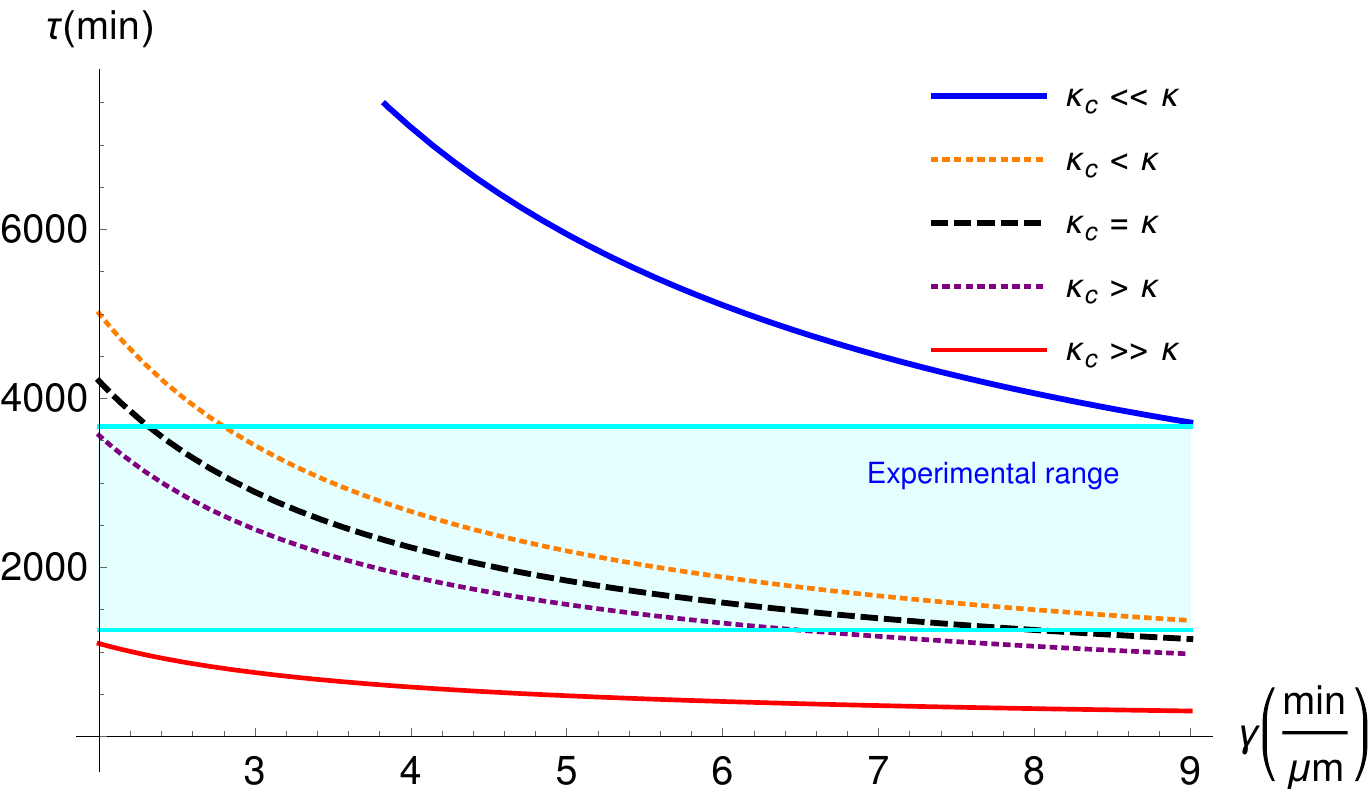}
 	  \caption{Characteristic release time of acetaminophen from erosive wax matrix as a function of the (undetermined) parameter $\gamma$, obtained from Eq.~(\ref{eq:tau}) with experimental values of capsule size, erosion rate and diffusion constant from~\cite{agata2011}.\label{fig6}} 
	\end{center}
\end{figure}

In this way, in Figure~\ref{fig6} the characteristic release time $\tau$ is shown as a function of $\gamma$ ($\mu \text{m/min})$, using five different values of $k_c$. The filled area between the vertical lines correspond to the reasonable experimental range values expected for $\tau$, ranging from around $63\%$ to $100\%$ of the release of acetaminophen from wax matrix in batch $5$~\cite{agata2011}. Five values of $k_c$ were chosen, $\{$4.5, 6.05, 6.37,  8.8, 9.5 $\} \times 10^{-2}\mu$m/min, corresponding to the cases where the erosion rate is slightly lower, slightly higher and equal to the crossover erosion rate, as well as much lower and much higher than the erosion rate of the setup ($6.37\times10^{-2}\mu$m/min). 

For $k_c = k$ (dashed line), the values of $\tau$ obtained from Eq.~(\ref{eq:tau}) are within the experimental range and its value is close to $63\%$ of release when $\gamma \approx 8~\mu\text{m/min}$. In order to compare this value of $\gamma$ to our computational model we consider the product $\gamma k_c$ which was found to be $\gamma k_c \approx 0.24$ in our model and approximately $0.5$ for experimental batch considered here.

The same figure also shows curves of $\tau$ in terms of $\gamma$ for $k_c < k$ and $k_c > k$, where we choose $\kappa_c$ as being displaced $5\%$ below or above $\kappa$ in each case. These curves behave similarly, showing that, within the current approach, batch $5$ is close to crossover erosion rate. In the cases where either $k_c << k$ (diffusion dominant, $k_c  = 4.5\times10^{-2}\mu$m/min) or $k_c >> k$ (erosion dominant, $k_c  = 9.5\times10^{-2}\mu$m/min)), $\tau$ is out of the experimental range, as shown in Fig.~\ref{fig6}. This is another evidence indicating that the release process is not determined by only one mechanism (either erosion or diffusion), and that, even without the AMCE compound erosion is relevant to the drug release mechanism.

%%+++++++++++++++++++++++++++++++++++++++++++++++++++++++++++++++++++++++   
\section{\label{sec:conclusion}Conclusion}
%%+++++++++++++++++++++++++++++++++++++++++++++++++++++++++++++++++++++++   

In this work we have extended the stochastic lattice model implemented in previous work from our group~\cite{gomesfilho2016, gomesfilho2020} 
to investigate the effects of membrane polymer erosion on the drug release mechanism. The Weibull function was found to describe with  good approximation the entire release curve in our model, for different erosion rates.  By comparing the time evolution of  membrane content and the amount of drug within the capsule we found that at $b_c\approx 1$ there is crossover between the dominant mechanisms: for $b$ values bigger than $b_c$, erosion is the predominant mechanism controlling the drug release while for $b < 1$ diffusion is governing the release. As shown in Fig.~\ref{fig2}, within each of this two regions the values of $b$ can be approximated by scaling laws of erosion rates with different exponents. 

We have also identified that within the crossover the characteristic time satisfy $\tau_c \approx L^{3/2}$, which can be justified through the Arrhenius relation for diffusion~(see Appendix~\ref{append:a}). The arguments used to demonstrate the size dependence on the characteristic time $\tau$ allowed us to propose the phenomenological function in~(\ref{eq:tau}), which nicely adjusted the drug release from the Monte Carlo simulations of the model proposed in this work. Experimental data for the release of acetaminophen from wax matrix was investigated using the expression for $\tau \equiv \tau(L,\kappa)$ and it was shown that erosion rates of the investigated devices (batch 5 of paper~\cite{agata2011}) were close to crossover erosion rate,  with the parameter $\gamma$ compatible with those obtained in our simulations. 

The computational results and the analysis of experimental data for acetaminophen release with Eq.~(\ref{eq:tau}) suggest that it should be possible to use scaling laws to improve our comprehension on the competition between different mechanisms governing the drug release dynamics. In particular, for the case of acetaminophen in wax matrix, in which the erosion can be controlled with addition of AMCE, our work indicate that through controlled increase of erosive agents, it should be possible to test the hypothesis in our models, as well as the usage of Eq.~(\ref{eq:tau}), or the small-$\kappa$ expression in~(\ref{eq:tau-low-k-exp}) to obtain the crossover erosion rate $\kappa_c$. 

Although the results presented in the current work started assuming linearly decaying membranes, it could also be generalized to exponentially decaying membranes with distribution probabilities satisfying a Poisson process~\cite{gopferich1993}. Even though outside the scope of the current work, these modifications should be straightforward to be performed, and we expect it should be compatible with most of our results, as can be readily inferred from Eq.~(\ref{eq:probability-membrane}).

It is important to stress that our model does not address the issue of indicating what drives the molecular relation between erosion and capsule size, geometry, drug load, and concentration of erosive compounds. 
Further theoretical or experimental data are needed for determining $\kappa$ and other parameters needed to calculate $\kappa_c$. Nevertheless, the novelty in our work is the proposition that in regimes where different mechanisms are controlling the drug release different scaling laws should be used to describe the relevant parameters describing the release process and that, by knowing the release behavior in those extreme regimes it should be possible to predict scaling behavior in the crossover. We expect that these ideas could be useful in providing some guidance to select the most useful properties and characteristics of capsules and devices for testing drug delivery systems.

\begin{figure}[tbp]
	\begin{center}
 	  \includegraphics[width=1\columnwidth]{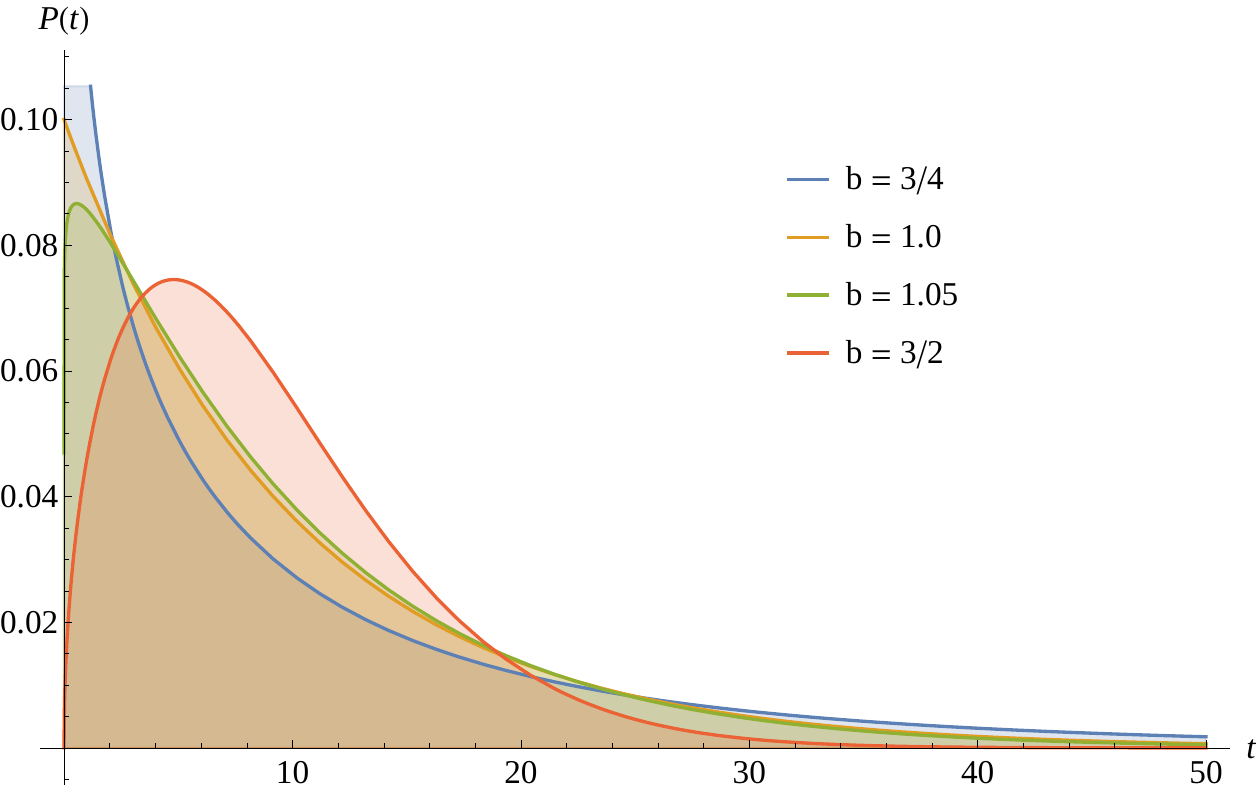}	 
	\caption{\label{fig:pdf} Plots of the  Weibull probability density distribution $P(t)$ against time $t$ for different values of $b$ (y-axis values in Fig.~\ref{fig2}). The time $t$ and $\tau = 10$ are in arbitrary units.}
	\end{center}
\end{figure}

\textit{Last remarks.\textemdash}
Finally, we would like to provide some arguments why Weibull distribution seems to correctly describe the drug release process. First, let us consider the drug released fraction, $R(t)$,   (\textit{e.g}, see Fig.~\ref{fig1}), which can be recognized as the Cumulative Probability Distribution (CPD)  while $P(t) = \frac{d R(t)}{dt}$ corresponds to the Weibull  Probability Density Distribution (PDF) with $t>0$, $b\geq0$ and $\tau>0$, which is also called as ``failure density" and gives the chance of a unit (drug particle) to fail (release) at time $t$~\cite{book:weibull,Corsaro21}.
For example, in Figure~\ref{fig:pdf} we show $P(t)$ against $t$ for different values of $b$. As we can see small deviations in  $b$, as those presented in Fig.~\ref{fig2}, changes the shape of the PDF, essentially altering the  characteristics  release times, \textit{e.g} the first moment of $P(t)$ (average time), $\langle t \rangle = \int_0^\infty t P(t)dt$, with $P(t)$ normalized. 

There is also a microscopic time associated with the particle trajectories (the time in which the drug particle takes to be released (fail)), which can be shorter or longer due to many factors, for example excluded volume interactions and  initial position of the particles  (whether or not it is closer to a leaking site, see Fig.~\ref{model}). 
For different realizations and particles, we must have  different trajectories, for a  capsule with mesoscopic size, we expect that the particle takes a very long time to reach a pore and be released. In other words, the microscopic release time is a rare event, and should satisfies the principles of the Extreme Value Distribution~\cite{Hansen20}, being directed related to  Mean-First-Passage time of a Brownian particle~\cite{redner01}, which may are the reasons for the Weibull Distribution to describe the drug release processes correctly, and it will investigate further in our future works.

% }
%%%+++++++++++++++++++++++++++++++++++++++++++++++++++++++++++++++++++++++++++++++++++++++++++++++++++++++++++++++++++++++++
\section*{Acknowledgments} 
%%%+++++++++++++++++++++++++++++++++++++++++++++++++++++++++++++++++++++++++++++++++++++++++++++++++++++++++++++++++++++++++

The authors acknowledge useful discussions with T. J. S. Carvalho,  O. T. Aranda and A. F. C. Campos. This study was financed by the agencies Coordena\c c\~ao de Aperfei\c coamento de Pessoal de N\'ivel Superior - Brasil (CAPES) - Finance Code 001, Conselho Nacional de Desenvolvimento Cient\'{i}fico e Tecnol\'{o}gico (CNPq) Grant No. CNPq-312497/2018-0 and
Funda\c{c}\~ao de Apoio a Pesquisa do Distrito Federal (FAPDF) Grant Numbers 0193.001616/2017 and 00193-00000120/2019-79. MSGF acknowledges financial support from \-FAPESP\- (Grant \# 2020/09011-9).

%193.000.581/\textcolor{red}{XXXX}.

%%%+++++++++++++++++++++++++++++++++++++++++++++++++++++++++++++++++++++++++++++++++++++++++++++++++++++++++++++++++++++++++
 \appendix
 \section{Derivation of crossover exponent $\mu_c$\label{append:a}}
%%%+++++++++++++++++++++++++++++++++++++++++++++++++++++++++++++++++++++++++++++++++++++++++++++++++++++++++++++++++++++++++
\noindent Let us start assuming that drug particles are leaving the capsule through an Arrhenius process in which the diffusion coefficient satisfies: 
\begin{equation}
    D \propto \exp{\left ( -\beta \Delta E \right )},
\end{equation}
where $\beta=1/k_B T$ with $k_B$ being the Boltzmann constant and $\Delta E$  the activation energy for the process of a drug particle to leave the capsule. For a purely diffusive system dimensional  analysis can be used to express the $D$ as:
\begin{equation}
D = \text{const.} \times \frac{l^2}{\tau},
\end{equation}
where $l$ is the capsule size, $\tau$ is the characteristic release time defined above and $\text{const.}$ is a constant which is possibly dependent on the system dimension and geometry. With these definitions  $\tau$ can be written as:
\begin{equation}
    \tau = A e^{\beta \Delta E},\label{eq:tau-arrhenius-basic}
\end{equation}
where  $A$ is a constant dependent on the system details. In order to consider the effect of erosion we will introduce a dependence on the erosion constant on all major functions, redefining~(\ref{eq:tau-arrhenius-basic}) as:
\begin{equation}
    \tau_\kappa = A_\kappa e^{\beta \Delta E_\kappa}.\label{eq:arrhenius}
\end{equation}

Besides this, we use the fact that $\tau_\kappa \equiv \tau(\kappa)$ depends on $\kappa$ through the scaling law Eq.~(\ref{eq:tau-a-mu}), and by using~(\ref{eq:arrhenius}) back in this expression one obtains the {\it activation free energy} for the drug release process for a certain erosion rate $\kappa$ as:
\begin{equation}
    \beta \Delta E_k  = \left [ 2 - \mu_\kappa \right ] \ln L - \ln \left [ \dfrac{2d D_\kappa  A_\kappa}{ l_0^2}  \right ],
\end{equation}
where $D_\kappa = D(\kappa)$ and $\mu_\kappa = \mu(\kappa)$.

From this latter expression, the simplest assumption about the value of crossover activation energy is that it could be the average between the two independent processes, $\Delta E_{c} = (\Delta E_D + \Delta E_E )/2$, where sub-indexes $c$, $E$, and $D$ denote the erosion constant values corresponding to \textit{crossover}, erosion controlled, and diffusion controlled systems. Through this latter assumption it is possible to calculate:
\begin{equation}
    \beta \Delta E_{c}   =  [2- \mu_c]\ln L - \ln \left ( \dfrac{2dD_{c} A_c}{l_0^2} \right),
\end{equation}
where
\begin{eqnarray}
   \mu_c  & = & \mu({ \kappa_c}) =  \frac{\mu_E + \mu_D}{2} = \frac{1}{2}, \label{eq:muc} \\
   D_c    & = &  \sqrt{D_E D_D}, \label{eq:Dc} \\
   A_c    & = &  \sqrt{A_E A_D}, \label{eq:Ac}
\end{eqnarray}
which provides a stronger derivation for the phenomenological value of the crossover scaling exponent $\mu_c$, as discussed on more phenomenological grounds in Sec.~(\ref{sec:tau-kappa}). It is also interesting to note that the diffusion~(\ref{eq:Dc}) and the coefficient $A_c$~(\ref{eq:Ac}) at the crossover satisfy a combination rule similar to that of Lorentz and Berthelot~\cite{Lorentz81,Berthelot98} for the Lennard-Jones interaction coefficients between atoms of different species.

%---------------------------------------------------%
\section{Low-$\kappa$ expression for $\tau$\label{append:b}}
%---------------------------------------------------%
\noindent Before proceeding let us define $x=\kappa/\kappa_c$, which will be used to express the characteristic time in Eq.~(\ref{eq:tau}) as:
\begin{equation}
    \dfrac{\tau (L,x)}{\tau_D} =  \dfrac{(l/l_0)^{ -(1 +x)^{-1}}}{1 - \exp(-\gamma \kappa_c x )}. \label{eq:tau2}
\end{equation}
If the erosion mechanism is dominant the system will be far from the \textit{crossover}, {\it i.e.}, $x \ll 1$. Within this limit Taylor expansions can be used to write:
\begin{equation}
    \left ( \frac{l}{l_0} \right )^{-\dfrac{1}{1+x}} \approx \left ( \frac{l}{l_0} \right )^{-1}  \left ( \frac{l}{l_0} \right )^x, \label{eq:low-k-num}
\end{equation}
and
\begin{equation}
    1 - \exp(-\gamma \kappa_c x ) \approx \gamma \kappa_c x, \label{eq:low-k-den}
\end{equation}
where it was implicit that $\gamma \kappa_c$ is a small number, as in our simulations. By using~(\ref{eq:low-k-num}) and~(\ref{eq:low-k-den}) in~(\ref{eq:tau2}) one gets:
\begin{equation}
     \dfrac{\tau (L,x)}{\tau_D} \approx \left ( \dfrac{l}{l_0} \right)^{-1}  \dfrac{L^x}{\gamma \kappa_c x} = \left ( \dfrac{l}{l_0} \right )^{-1} \dfrac{\exp(x \ln L)}{\gamma \kappa_c x}.
\end{equation}
If one assumes that the capsule is small, in the sense that $x < x\ln L \ll 1$, it is possible to simplify this equation even further by writing  it as:
\begin{equation}
    \tau (L,x) \approx  \left (\dfrac{\tau_D }{ \gamma \kappa_c} \right) \left ( \dfrac{l}{l_0} \right)^{-1} \left ( \dfrac{1}{x} + \ln  \dfrac{l}{l_0} \right ). \label{eq:tau-low-k-exp}
\end{equation}

Note that the right hand side of this expression was rearranged to emphasize that the contributions to $\tau$ are coming from three different terms. The characteristic release time $\tau_D$ is modulated by $\gamma \kappa_c$ because this is the low limit of function $F(\kappa)$ defined in~(\ref{eq:f}) to reduce the diffusion constant inside the capsule as an effect of the presence of the erosive membrane covering it. The term $(l/l_0)^{-1}$ appears due to the choice made in Eq.~(\ref{eq:tau-a-mu}), which is associated to the linear scaling linear law dependence of $\tau$ with size in the slow erosive regime, while the last term express that $\tau$ increases with $1/x$ and that there is a logarithmic correction with the capsule size, that is independent of $\kappa$.

It is interesting to rewrite Eq.~(\ref{eq:tau-low-k-exp}) as:
\begin{equation}
    L \dfrac{\tau}{\tau_D} = A \dfrac{1}{\kappa} + B,
\end{equation}
where $A=1/\gamma$ and $B=\ln L / (\gamma \kappa_c)$, which is amenable to simple linear interpolation in terms of $1/\kappa$. In this way, experimental values of $\gamma$ and $\kappa_c$ could be obtained through $\gamma = A^{-1}$ and $\kappa_c = \ln L A/B$.

\bibliography{references}

% %%%+++++++++++++++++++++++++++++++++++++++++++++++++++++++++++++++++++++++++++++++++++++++++++++++++++++++++++++++++++++++++
% \nomenclature{$N(t)$}{Number of drug particles as a function of time.}
% \nomenclature{$N_0$}{Initial number of drug particles.}
% \nomenclature{$t$}{Time in units of Monte Carlo Steps (MCS).}
% \nomenclature{$b$}{The Weibull  release parameter.}
% \nomenclature{$\tau$}{Characteristic release time .}

% \nomenclature{$L$}{Size (number of lattice spaces).}
% \nomenclature{$l_0$}{pore length [unit of length].}
% \nomenclature{$l = Ll_0$}{ capsule size [unit of length].}
% \nomenclature{$M_0$}{Initial number of membrane sites.}
% \nomenclature{$M(t)$}{Average number of membrane sites.}
% \nomenclature{$\kappa$}{Erosion rate constant [unit of length]/[unit of time]}

\end{document}